\documentclass[a4paper,10pt]{article}
\pdfoutput=1 
\usepackage{jheppub}
\usepackage[T1]{fontenc} 
\usepackage{float}
\usepackage{tikz-cd}
\usepackage{braket}
\usetikzlibrary{cd}
\definecolor{rindou1}{rgb}{0.4431,0.2862,0.7960}
\definecolor{rindou2}{rgb}{0.0078,0.1215,0.4392}
\definecolor{lapis}{rgb}{0.0.0470,0.2941,0.5568}
\definecolor{emerald}{rgb}{0.31, 0.78, 0.47}
\definecolor{deepsaffron}{rgb}{1.0, 0.6, 0.2}
\definecolor{pinegreen}{rgb}{0.0, 0.47, 0.44}
\definecolor{majorelleblue}{rgb}{0.38, 0.31, 0.86}
\definecolor{jade}{rgb}{0.0, 0.66, 0.42}
\definecolor{teal}{rgb}{0.0, 0.5, 0.5}
\definecolor{darkcyan}{rgb}{0.0, 0.55, 0.55}
\definecolor{jazzberryjam}{rgb}{0.65, 0.04, 0.37}
\definecolor{electricviolet}{rgb}{0.56, 0.0, 1.0}
\definecolor{regalia}{rgb}{0.32, 0.18, 0.5}
\definecolor{burgundy}{rgb}{0.5, 0.0, 0.13}
\definecolor{indigo(web)}{rgb}{0.29, 0.0, 0.51}
\definecolor{cerise}{rgb}{0.87, 0.19, 0.39}
\definecolor{darkbyzantium}{rgb}{0.36, 0.22, 0.33}
\definecolor{darkscarlet}{rgb}{0.34, 0.01, 0.1}
\definecolor{tyrianpurple}{rgb}{0.4, 0.01, 0.24}
\definecolor{bondiblue}{rgb}{0.0, 0.58, 0.71}
\definecolor{amaranth}{rgb}{0.9, 0.17, 0.31}
\definecolor{blue(munsell)}{rgb}{0.0, 0.5, 0.69}
\definecolor{yaleblue}{rgb}{0.06, 0.3, 0.57}
\definecolor{iris}{rgb}{0.35, 0.31, 0.81}
\definecolor{darkred}{rgb}{0.55, 0.0, 0.0}
\definecolor{slateblue}{rgb}{0.42, 0.35, 0.8}
\usepackage[T1]{fontenc}
\usepackage{datetime2}
\usepackage[english]{babel}
\usepackage[utf8]{inputenc}
\usepackage{graphicx}
\usepackage{color}
\usepackage{amsfonts,amsthm}
\usepackage{bm,bbm}
\usepackage{mathrsfs}
\usepackage{amsmath}
\usepackage{float}
\usepackage{pgf,tikz}
\usepackage{mathrsfs}
\usepackage{xcolor}
\usetikzlibrary{arrows}

\newcommand{\z}{\overline{z}}

\newcommand{\A}{\mathcal{A}}
\newcommand{\ab}[1]{\langle{#1\rangle}}

\title{\boldmath Gauge and Gravity Amplitudes on the Celestial Sphere}

\author{Nikhil Kalyanapuram}

\affiliation{Department of Physics and Institute for Gravitation and the Cosmos, The Pennsylvania State University, University Park PA 16802, USA}

\emailAdd{nkalyanapuram@psu.edu}

\abstract{The analytic structures of scattering amplitudes in gauge theory and gravity are examined on the celestial sphere. The celestial amplitudes in the two theories - computed by employing a regulated Mellin transform - can be compared at low multiplicity. It is established by direct computation that up to five external particles, the double copy relations of Kawai, Lewellen and Tye continue to hold identically, modulo certain multiplicative factors which are explicitly determined. Supersymmetric representations of the amplitudes are utilized throughout, manifesting the double copy structure between $\mathcal{N}=4$ super Yang-Mills and $\mathcal{N}=8$ supergravity on the celestial sphere.}

\begin{document} 
\maketitle
\flushbottom

\section{Introduction}
Scattering amplitudes of massless particles exhibit a number of properties - physical and mathematical - which make them rather interesting as objects of independent study. It has been observed in particular that certain special choices of bases or variables manifest these properties most explicitly.

A particular example of a simplifying choice of variables is the so-called spinor helicity framework. The spinor helicity variables manifest Lorentz invariance and little group scaling in a manner than makes the physical properties of scattering amplitudes in gauge theory and gravity essentially transparent. Notably, the properties of this notation are sufficient to constrain the three-particle amplitudes for massless particles of any spin more or less uniquely \cite{Cheung:2017pzi}. 

A basis which has enjoyed much recent interest is what has come to be known as the Mellin basis or the SL($2$,C) basis \cite{Pasterski:2016qvg,Pasterski:2017ylz}. When expressed using this basis, external states are boost eigenvectors, as opposed to the momentum eigenstates we usually employ. The use of the Mellin basis has been inspired largely by the hope of realizing a case of flat space holography. The argument is as follows.

Consider a scattering amplitude involving $n$ massless particles. Massless particles are characterized by null momenta. In particular, when a given particle $k$ has energy $\omega_{k}$, its four-momentum admits the decomposition

\begin{equation}
    p^{\mu}_{k} = \omega_{k}q^{\mu}_{k} 
\end{equation}
where

\begin{equation}
    q^{\mu}_{k} = \left(1+z_{k}\z_{k},z_{k}+\z_{k},-i(z_{k}-\z_{k}),1-z_{k}\z_{k}\right).
\end{equation}

The labels $z_{k}$ and $\z_{k}$ are to be regarded as fixed points on a celestial sphere isomorphic to $\mathbb{CP}^{1}$. In deference to tradition, we will refer to this system of coordinates as stereographic coordinates. 

Clever (and judicious) uses of stereographic coordinates have been quite helpful in studying various boundary aspects of the flat-space $S$-matrix for massless particles. More concretely, among other things there has been a resurgence of interest in the various implications of what are now called \emph{asymptotic symmetries}. Asymptotic symmetries are global phenomena that are best studied as transformations on the celestial sphere. In the case of gauge theory, the group of asymptotic symmetries can be related to large gauge transformations. Invariance of the $S$-matrix under this group has been shown to imply the soft theorems for spin-$1$ emissions \cite{Strominger:2015bla,He:2014cra,He:2015zea,Strominger:2017zoo,Laddha:2017vfh,Campiglia:2015qka,Campiglia:2016hvg,Campiglia:2018dyi} at leading and subleading order.

More subtle is the case of scattering amplitudes in gravity. That radiative degrees of gravity are encoded delicately on null hypersurfaces at infinity in asymptotically flat spacetimes was first understood by Penrose \cite{Penrose:1965am} and by Newman and Penrose \cite{Newman:1968uj} in the 1960s. While the analyses carried out in these works made very clear the asymptotic behaviour of radiative fields for both spin-$1$ and spin-$2$ fields, a precise connection to scattering amplitudes relevant for quantum field theory remained more or less an open question. Progress was made when it was understood that the asymptotic behaviour of massless fields in gravity could be studied in light of a large symmetry group at infinity, now known as the Bondi-Metzner-Sachs (BMS) group \cite{PhysRev.128.2851,Bondi:1962px}. Specifically, it has been realized in recent years \cite{He:2014laa,Campiglia:2014yka,Campiglia:2016efb} that the $S$-matrix in gravity is invariance under the BMS group, the Ward identities of which are equivalent to the soft graviton theorem of Weinberg \cite{Weinberg:1965nx}. Of particular interest is that these observations suggest that the essential analytic behaviour of scattering amplitudes in gauge theory and gravity are in reality boundary phenomena. More to the point, this point of view has raised the hope that a genuine example of flat space holography may be in sight.

Conjecturally, such a duality would be developed by providing an explicit map between field operators $O_{i}$ having four-dimensional support in the bulk and operators $\mathcal{O}_{i}^{\Delta}$ in a putative conformal field theory, with conformal weights $\Delta$ specified. Additionally, these would have to specify a dual description of the $S$-matrix in the schematic manner,

\begin{equation}
    \bra{0}a(p_{1})\cdots a^{\dagger}(p_{n})\ket{0} \Longleftrightarrow \langle{\mathcal{O}(z_{1})^{\Delta_{1}}\cdots \mathcal{O}(z_{n})^{\Delta_{n}}\rangle}.
\end{equation}
While this is conjectural, a key step in developing such a duality would be to prescribe a map, perhaps in the form of a change of basis, that can be used to move between the two pictures. It is in this context that the Mellin basis makes its appearance. 

Suppose we have a scattering amplitude $\A_{n}$, where we keep the delta function imposing momentum conservation in place, rather than working with the stripped amplitude. While we point the reader to \cite{Pasterski:2016qvg,Pasterski:2017ylz} for the attendant technical arguments, it was observed that a motion into a conformal primary basis could be achieved for gluon amplitudes using the so-called Mellin transform, applied to the energies of the particles. Put quantitatively, the holographic $S$-matrix is defined implicitly on $n$ copies of the celestial sphere by transforming the energies according to

\begin{equation}
    \mathcal{M}_{n} = \int \prod_{i=1}^{n} \omega_{i}^{i\Lambda_{i}} d\omega_{i} \times \A_{n}.
\end{equation}
The $\Lambda_{i}$ are related to the conformal weights $h,\overline{h}$ and helicities $\delta_{i}$ according to

\begin{equation}
    h_{i} = \frac{1 + i\Lambda_{i}-\delta_{i}}{2},
\end{equation}
and

\begin{equation}
    \overline{h}_{i} = \frac{1 + i\Lambda_{i}+\delta_{i}}{2}.
\end{equation}
We point out that the above symbols match the (unfortunate) convention in the literature. In particular, it should be noted that $\overline{h}_{i}$ does not indicate a complex conjugation.

Although this map is very robust for gluon amplitudes, which are to our best knowledge the final story, one faces a more delicate situation with gravitational amplitudes. It is clear from the definition of the Mellin transform that the amplitude must be known at arbitrarily large energies. However, gravity amplitudes have very poor ultraviolet behaviour, making the Mellin transform ill-defined. Curing this problem requires either a consistent understanding of gravity in the ultraviolet (a tall order) or a regulated form of the transform, which corrects for the fact that gravitational amplitudes ultimately come from an effective field theory. This problem was studied in \cite{Banerjee:2018gce,Banerjee:2018tun}, where a regulated Mellin transform was proposed and in \cite{Banerjee:2019prz}, where it was applied to four-particle amplitudes in gravity and the conformal soft theorems were studied. 

The regulated Mellin transform is qualitatively different from the unregulated one. One important distinction is the fact that it maps the four dimensional theory onto a three-dimensional one, not a two-dimensional theory. The additional degree of freedom is furnished by the retarded times $u_{i}$, which appear according to

\begin{equation}
    \mathcal{M}_{n} = \int \prod \omega^{i\Lambda_{i}}_{i}e^{i\epsilon_{i}u_{i}\omega_{i}}d\omega_{i} \A_{n},
\end{equation}
where the $\epsilon_{i}$ are $+1$ for outgoing and $-1$ for incoming particles. Needless to say, the regulated Mellin transform is well-defined under conditions where

\begin{equation}
    \mathrm{sgn}(\mathrm{Im}u_{i}) = \epsilon_{i}.
\end{equation}
Although regulated Mellin amplitudes are somewhat more complicated than those which are not, they provide finite results for any amplitudes we choose, making it possible to study even effective field theories in a holographic context. With this background material, we are brought to the main goal of this paper, namely a systematic study of scattering amplitudes in gauge theory and gravity making use of the regulated Mellin transform. 

Recent years have seen progress in understanding formal relationships between scattering amplitudes in gauge theory and gravity. An interesting example of such relationships is referred to often by the umbrella term \emph{double copy}. Put simply, a double copy relation is one that realizes scattering amplitudes in one theory (say theory A) as a square of some sort of amplitudes in another theory (say theory B). While there are a number of avatars of this idea, the most concrete of these stems from a well known construction in string theory, known as the Kawai-Lewellen-Tye relation \cite{Kawai:1985xq}. In its simplest form, the KLT relation states that given a basis of $(n-3)!\times (n-3)!$ gauge theory amplitudes for $n$ gluons, labelled by two sets of $(n-3)!$ permutations $\lbrace{\alpha\rbrace}$ and $\lbrace{\beta\rbrace}$ of the external legs, the (stripped) gravity amplitude for $n$ particles (all at tree level) can be computed by the convolution of (stripped) gauge theory amplitudes

\begin{equation}
    \A_{n,grav} = \sum_{\alpha,\beta}\A_{n,gauge}[\alpha]K[\alpha|\beta]^{-1}\A_{n,gauge}[\beta].
\end{equation}
The matrix $K[\alpha|\beta]$, which has $(n-3)!\times (n-3)!$ represents scattering amplitudes of a theory known as the biadjoint scalar theory. Purely kinematic in character, the KLT relation recasts gravitational amplitudes as quite literally two copies of gauge theory amplitudes. 

Interesting as it is, the KLT relation hold strictly in momentum space. As we have just argued however, there are other bases which can be employed while studying scattering amplitudes that manifest other phenomena of interest. Accordingly, the extent to which the KLT relation holds away from momentum space is a worthwhile question to ponder.

In this article, one of our tasks will be to clarify the extent to which the KLT relation continues to apply when a basis change into Mellin space is carried out.  As we will see, it is by no means obvious that it should continue to hold, and indeed, it does not in general. More specifically, we will see that modulo trivial corrections, the KLT relation is still valid for low multiplicity ($\leq 5$ external particles). 

How we go about studying this problem is as follows. There is evidence to suggest that double copy relations are essentially descendants of analytic phenomena in string theory \cite{DHoker:1989cxq,Mizera:2019gea,Mizera:2019blq,Kalyanapuram:2021xow,Kalyanapuram:2021vjt}. Informed by this, in the present article we employ supersymmetric representations of gauge theory and gravity amplitudes, keeping track of $\mathcal{N}=4$ and $\mathcal{N}=8$ supersymmetries respectively using $C$ matrices\footnote{$C$ matrices encode the supersymmetric structure of the scattering amplitudes.}. As we shall see, these supersymmetric representations follow the path of KLT, namely that they no longer continue to hold na\"ively beyond five particle scattering. Indeed, both of these occurrences have a common origin, the ultimately mundane details of which we will clarify in section \ref{sec:3}.

In the following sections, we will provide complete analytic details on the computation of regulated Mellin space amplitudes up to five particle scattering. The cases of three-particle, four-particle and five-particles are dealt with consecutively in sections \ref{ssec:1}, \ref{ssec:2} and \ref{ssec:3} respectively. These are followed in \ref{sec:3} by a discussion of the results as well as the obstructions one would encounter at higher loops. The article concludes with a review of possible future directions in \ref{sec:4}.

\paragraph{A word on notation:}
Before getting into calculations, we need to set up the notation for spinor helicity variables. Consider two momenta $p_{i}$ and $p_{j}$. Inner products of momenta appear in the form $2\epsilon_{i}\epsilon_{j}p_{i}\cdot p_{j}$. To decompose this into spinor helcity brackets, we write

\begin{equation}
    2\epsilon_{i}\epsilon_{j}p_{i}\cdot p_{j}= \ab{ij}[ij].
\end{equation}
In this paper, we will deal with MHV amplitudes for simplicity, since the multiplicities in which we are interested have a single nontrivial helicity sector, namely the MHV sector. The others can be obtained by conjugation. This means that we will only work with angle brackets. To avoid a profusion of essentially irrelevant sign factors, we adopt the convention

\begin{equation}
    \langle{ij\rangle} = \sqrt{\omega_{i}\omega_{j}}(z_{i}-z_{j})
\end{equation}
\begin{equation}
    [ij] = \epsilon_{i}\epsilon_{j}\sqrt{\omega_{i}\omega_{j}}(\z_{i}-\z_{j})
\end{equation}
where the energies $\omega_{i}$ are real and positive. The choice is entirely arbitrary, but we choose it in this manner to make the expressions to come less cryptic. We point out that this is the opposite of the convention used in \cite{Pasterski:2017ylz}. 

It is worth making a note of the profusion of notation that will be met by the reader in the following section. Throughout, we will use $u_{i}$ to denote retarded times, which are always understood to have imaginary parts $\epsilon_{i}\delta$, for positive $\delta$. $z_{ij}$ will be used to denote the difference $z_i-z_j$, with analogous use being made of $\z_{ij}$. Finally, $\Lambda_i$ always denote the weights appearing in the Mellin transform. 

\paragraph{Note added:}
When this paper was nearing completion, the paper \cite{Casali:2020uvr} appeared on the arXiv, which has some overlap with this paper in content, although the spirit of analysis in this work is different. In \cite{Casali:2020uvr}, the authors recast celestial amplitudes (unregulated) in a CHY form and derive BCJ-like dualities using twisted cohomology. In this work however, we have sought to find the limit of applying KLT with little or no change to celestial amplitudes and study the preservation of manifest supersymmetry. Comparing and contrasting these two properties has been our goal here, while Casali and Sharma have focused on deriving a more concrete realization of BCJ.

\section{Regulated Mellin Amplitudes at Low Multiplicity}\label{sec:2}

\subsection{Three-Particle Amplitudes}\label{ssec:1}
The three particle amplitudes for massless interacting particles are fixed completely by conditions of Lorentz invariance and little group scaling. Indeed, when put in spinor helicity notation, the double copy prescription is quite literal, taking the form,

\begin{equation}
    A_{3}(2s_{1},2s_{2},2s_{3}) = A_{3}(s_{1},s_{2},s_{3})^{2}.
\end{equation}
While we refer the interested reader to \cite{Cheung:2017pzi} for a complete treatment of the details involved in constructing three particle amplitudes, it is worth pointing out that this fact is kinematically guaranteed when working in spinor helicity variables, which make manifest little group scaling. The question we would like to understand and answer in this section is to what extent this squaring procedure is valid when handling three particle amplitudes in the Mellin basis. As we shall see, insofar as we are dealing with three particles, the squaring procedure is all but preserved, with minor deviations.

Since we will only be dealing with amplitudes in gauge theory and gravity, our first task will be to present these in a form that is convenient to use for explicit calculations. In particular, we will employ the following supersymmetric form of the amplitudes 

\begin{equation}
    \mathcal{A}_{3}\left(123\right) = \frac{\delta^{4}\left(\sum_{i}\epsilon_{i}p_{i}\right)\delta^{8}\left(C(\lambda)\right)}{\ab{12}\ab{23}\ab{31}},
\end{equation}

\begin{equation}
    \mathcal{A}_{3}\left(123\right) = \frac{\delta^{4}\left(\sum_{i}\epsilon_{i}p_{i}\right)\delta^{16}\left(C(\lambda)\right)}{(\ab{12}\ab{23}\ab{31})^{2}}.
\end{equation}

where we have denoted

\begin{equation}
    C(\lambda)_{\alpha}^{I} = \sum_{i=1}^{3}\lambda^{(i)}_{\alpha}\eta^{I}_{(i)}.
\end{equation}
$I$ is an index taking values from $1$ to $4$ for gauge theory and from $1$ to $8$ for gravity, indicating the number of supersymmetries.

Before getting into the more cumbersome details of the Mellin transform, which replaces the energies $\omega_{i}$ by so-called weights $\Lambda_{i}$, we can get a sense for how the transform behaves by a change of variables, introducing a proper time parameter $s$ and simplex variables $\sigma_{i}$ according to

\begin{equation}
    \omega_{i} = s\sigma_{i}.
\end{equation}
The $\sigma_{i}$ are called simplex variables as they are constrained by the relation

\begin{equation}
    \sum_{i}\sigma_{i} = 1.
\end{equation}

This change of variables for small multiplicity trivializes the simplex integrals due to a residue theorem, while the $s$ integrals can be evaluated independently. Accordingly, it is valuable to extract out the $s$ dependence from the foregoing amplitudes, which is homogeneous for the cases with which we will be concerned. For the gauge theory case we have

\begin{equation}
\mathcal{A}_{3}\left(123\right) = \frac{s^{-4}\times s^{4}}{s^{3}} \hat{\mathcal{A}}_{3}\left(123\right)
\end{equation}
and 

\begin{equation}
\mathcal{A}_{3}\left(123\right) = \frac{s^{-4}\times s^{8}}{s^{6}} \hat{\mathcal{A}}_{3}\left(123\right)
\end{equation}
where the hat (here and more generally) indicates factorization of the $s$ part outside the amplitude. Every variable in $\hat{\mathcal{A}}$ depends only on the simplex variables and positions $z_{i}$ on the celestial sphere.

It is instructive to qualify the powers of $s$ that appear in the numerators, since it is a trick we will employ at various points in this article. The gauge theory case will suffice for this purpose. The $s^{-4}$ arises due to extracting the homogeneous dependence of the momentum conserving delta function on $s$. The power $s^4$ on the other hand comes from the supersymmetric delta function. Analogously, due to twice the number of supersymmetries involved, we have a corresponding factor of $s^8$ in the supergravity case. We can now evaluate the Mellin space integrals in turn. Starting with the $s$ integration we have

\begin{equation}
    \int s^{3-1 + i\sum_{i}\Lambda_{i} -3} e^{is\sum_{i}\epsilon_{i}\sigma_{i}u_{i}} ds
\end{equation}
for the gauge theory case and 

\begin{equation}
    \int s^{3-1 + i\sum_{i}\Lambda_{i} -2} e^{is\sum_{i}\epsilon_{i}\sigma_{i}u_{i}} ds
\end{equation}
for the gravity amplitude. These evaluate to

\begin{equation}
    \frac{\Gamma\left[i\sum_{i}\Lambda_{i}\right]}{\left(i\sum_{i}\epsilon_{i}\sigma_{i}u_{i}\right)^{i\sum_{i}\Lambda_{i}}}
\end{equation}
and

\begin{equation}
    \frac{\Gamma\left[1+i\sum_{i}\Lambda_{i}\right]}{\left(i\sum_{i}\epsilon_{i}\sigma_{i}u_{i}\right)^{1+i\sum_{i}\Lambda_{i}}}
\end{equation}
respectively.

The evaluation of the simplex integrals has been done on several occasions \cite{Pasterski:2016qvg,Pasterski:2017ylz} already, but it's instructive to repeat them. The decisive part of the computation is the transformation of the delta function constraints

\begin{equation}
    \delta^{4}\left(\sum_{i}\sigma_{i}\epsilon_{i}q_{i}\right)\delta\left(1-\sum_{i}\sigma_{i}\right) = R_{3}C_{3}\prod_{i}\delta(\sigma_{i} - \sigma^{*}_{i}).
\end{equation}
where \cite{Banerjee:2019prz}

\begin{equation}
\begin{aligned}
     C_{3} =& \frac{1}{\sigma^{*}_{1}\sigma^{*}_{2}\sigma^{*}_{3}D_3^2}\\
     R_{3} = & \delta(\z_{12})\delta(\z_{13})
\end{aligned}
\end{equation}
and

\begin{equation}
\begin{aligned}
     \sigma^{*}_{1} =& \frac{z_{23}}{D_{3}}\\
     \sigma^{*}_{2} = & -\frac{\epsilon_{1}\epsilon_{2}z_{13}}{D_{3}}\\
     \sigma^{*}_{3} =& \frac{\epsilon_{1}\epsilon_{3}z_{12}}{D_{3}}& 
\end{aligned}
\end{equation}
where

\begin{equation}
    D_{3} = (1-\epsilon_{1}\epsilon_{2})z_{13} + (\epsilon_{1}\epsilon_{3}-1)z_{12}.
\end{equation}
Given this transformation, the simplex integrals reduce to residues on the $\sigma^{*}_{i}$, giving us

\begin{equation}
  \mathcal{M}_{3,gauge} =  R_{3}C_{3} \frac{\Gamma\left[i\sum_{i}\Lambda_{i}\right]}{\left(i\sum_{i}\epsilon_{i}\sigma^{*}_{i}u_{i}\right)^{i\sum_{i}\Lambda_{i}}}\frac{\delta^{8}\left(\hat{C}(\lambda^{*})\right)}{\sigma^{*}_{1}\sigma^{*}_{2}\sigma^{*}_{3}z_{12}z_{23}z_{31}}\prod_{i}(\sigma^{*}_{i})^{i\Lambda_{i}}
\end{equation}
for gauge theory and 

\begin{equation}
 \mathcal{M}_{3,grav} = R_{3}C_{3} \frac{\Gamma\left[1+i\sum_{i}\Lambda_{i}\right]}{\left(i\sum_{i}\epsilon_{i}\sigma^{*}_{i}u_{i}\right)^{1+i\sum_{i}\Lambda_{i}}}\frac{\delta^{16}\left(\hat{C}(\lambda^{*})\right)}{(\sigma^{*}_{1}\sigma^{*}_{2}\sigma^{*}_{3}z_{12}z_{23}z_{31})^{2}}\prod_{i}(\sigma^{*}_{i})^{i\Lambda_{i}}
\end{equation}
for gravity. We remark that $\hat{C}(\lambda^{*})$ has been used to indicate the supersymmetric $C$ matrix with the spinor helicity variables evaluated at the localized values of the simplex variables. In this form, it is evident that the double copy is maintained, at least qualitatively. To see how this relates to the KLT relations, we observe that the KLT kernel for three particles, enforcing momentum conservation is simply

\begin{equation}
    \delta^{4}\left(p_{1}+p_{2}+p_{3}\right).
\end{equation}

It is a simple exercise to verify that the modified Mellin transform of this quantity is just

\begin{equation}
   \mathcal{K}_{3} =  R_{3}C_{3}\frac{\Gamma\left[-1+i\sum_{i}\Lambda_{i}\right]}{\left(i\sum_{i}\epsilon\sigma^{*}_{i}u_{i}\right)^{-1+i\sum_{i}\Lambda_{i}}}\prod_{i}(\sigma^{*}_{i})^{i\Lambda_{i}}.
\end{equation}

From direct inspection, we see that the KLT relation continues to hold, other than a prefactor which is readily computed,

\begin{equation}
    \Delta_{3} = \frac{\Gamma\left[1+i\sum_{i}\Lambda_{i}\right]\Gamma\left[-1+i\sum_{i}\Lambda_{i}\right]}{\Gamma\left[i\sum_{i}\Lambda_{i}\right]^{2}}.
\end{equation}
This definition allows us to state schematically,

\begin{equation}
    \mathcal{M}_{3,grav} =\mathcal{M}_{3,gauge} \Delta_{3}\mathcal{K}_{3}^{-1} \mathcal{M}_{3,gauge}.
\end{equation}
In other words, the KLT relation is essentially preserved, with a correction term having no dependence on the kinematics\footnote{It was pointed out to the author in a private communication that this could be reabsorbed into a redefinition of the gauge theory and gravity coupling constants. This happens to not be the case however, since the correction term is dependent on the number of particles. Accordingly, they do not amount to an overall constant multiplying the amplitude which can be renormalized away.}. This fact provides a valuable hint as to the conditions under which we expect continued applicability of relations of KLT type. When the Mellin transform is carried out, as we shall see, on amplitudes at low multiplicity, the essential kinematical structure of these amplitudes is kept intact. Under these circumstances, relations such as the double copy will be preserved, with non-kinematical corrections. Keeping this in mind, we can look at more complicated examples to see how this argument plays out.

\subsection{Four-Particle Amplitudes}\label{ssec:2}
We shall now be concerned with generalizing the analysis of the last section to four particle scattering amplitudes in gauge theory and gravity. We start with the unique four particle scattering amplitude for spin $1$ particles,

\begin{equation}
    \A_{4}\left(1234\right) = \frac{\delta^{8}\left(C(\lambda)\right)\delta^{4}\left(\sum_{i}p_{i}\right)}{\ab{12}\ab{23}\ab{34}\ab{41}}
\end{equation}
where

\begin{equation}
    C(\lambda)_{\alpha}^{I} = \sum_{i=1}^{4}\lambda^{(i)}_{\alpha}\eta^{I}_{(i)}
\end{equation}
has been expressed in terms of the spinor helicity variables and Grassmann supersymmetry generators. We have the modified Mellin amplitude

\begin{equation}
    \mathcal{M}_{4,gauge} = \int \prod_{i}\omega^{i\Lambda_{i}}_{i} \A_{4}\left(1234\right)e^{i\sum_{i}\epsilon_{i}\omega_{i}u_{i}}d\omega_{i}
\end{equation}
where it is to be kept in mind that 

\begin{equation}
    \mathrm{sgn}(\mathrm{Im}u_{i}) = \epsilon_{i}.
\end{equation}

The change of variables defined by

\begin{equation}
    \sigma_{i} = s^{-1}\omega_{i}
\end{equation}
and

\begin{equation}
    \sum_{i}\sigma_{i} = 1
\end{equation}
transforms the Mellin space amplitude into

\begin{equation}
    \mathcal{M}_{4,gauge} = \int s^{4-1+i\sum_{i}\Lambda_{i}}\int \prod_{i}\sigma^{i\Lambda_{i}}_{i} \A_{4}\left(1234\right)e^{is\sum_{i}\epsilon_{i}\sigma_{i}u_{i}}ds d\sigma_{i}.
\end{equation}
The integral over $s$ is evaluated once again by noting the homogeneous scaling of the amplitude given by

\begin{equation}
    \mathcal{A}_{4}\left(1234\right) = s^{-4}\hat{\A}_{4}\left(1234\right)
\end{equation}
where as usual the hat has been used to indicate independence from the variable $s$. Performing the integral over $s$ we obtain

\begin{equation}\label{eq:3.9}
    \mathcal{M}_{4,gauge} = \frac{\Gamma\left[i\sum_{i}\Lambda_{i}\right]}{\left(i\sum_{i}\epsilon_{i}\sigma_{i}u_{i}\right)^{i\sum_{i}\Lambda_{i}}}\int \prod_{i}\sigma^{i\Lambda_{i}}_{i} \A_{4}\left(1234\right) d\sigma_{i}.
\end{equation}

The $\sigma_{i}$ integrals are localized by the five delta functions constraining the momenta and simplex. They are effected by transforming the delta functions according to \cite{Banerjee:2019prz}

\begin{equation}\label{eq:3.10}
    \delta^{4}\left(\sum_{i}\epsilon_{i}\sigma_{i}q_{i}\right) \delta\left(1-\sum_{i}\sigma_{i}\right) = C_{4}R_{4}\prod_{i}\delta(\sigma_{i}-\sigma^{*}_{i})
\end{equation}
where

\begin{equation}
    \begin{aligned}
    C_{4} &= \frac{1}{4}\\
    R_{4} &= \delta\left(|z_{12}z_{34}\z_{13}\z_{24}-\z_{12}\z_{34}z_{13}z_{24}|\right)
    \end{aligned}
\end{equation}
and

\begin{equation}
    \begin{aligned}
    \sigma^{*}_{1} &= \frac{\epsilon_{1}\epsilon_{4}}{D_{4}}\frac{z_{24}\z_{34}}{z_{12}\z_{13}}\\
    \sigma^{*}_{2} &= \frac{\epsilon_{2}\epsilon_{4}}{D_{4}}\frac{z_{34}\z_{14}}{z_{23}\z_{12}}\\
    \sigma^{*}_{3} &=\frac{\epsilon_{3}\epsilon_{4}}{D_{4}}\frac{z_{24}\z_{14}}{z_{23}\z_{13}}\\
    \sigma^{*}_{4} &= \frac{1}{D_{4}}
    \end{aligned}
\end{equation}
where we have denoted

\begin{equation}
    D_{4} = \frac{(1-\epsilon_{1}\epsilon_{4})}{D_{4}}\frac{z_{24}\z_{34}}{z_{12}\z_{13}} - \frac{(1-\epsilon_{2}\epsilon_{4})}{D_{4}}\frac{z_{34}\z_{14}}{z_{23}\z_{12}} + \frac{(1-\epsilon_{3}\epsilon_{4})}{D_{4}}\frac{z_{24}\z_{14}}{z_{23}\z_{13}}.
\end{equation}

The delta functions in (\emph{\ref{eq:3.10}}) determine the integral (\emph{\ref{eq:3.9}}) by fixing the simplex variables, giving

\begin{equation}\label{eq:3.14}
   \mathcal{M}_{4,gauge}(1234) =  \frac{\Gamma\left[i\sum_{i}\Lambda_{i}\right]}{\left(i\sum_{i}\epsilon_{i}\sigma_{i}u_{i}\right)^{i\sum_{i}\Lambda_{i}}}C_{4}R_{4}\frac{\delta^{8}\left(C(\sigma^{*})\right)}{\sigma^{*}_{1}\sigma^{*}_{2}\sigma^{*}_{3}\sigma^{*}_{4}z_{12}z_{23}z_{34}z_{41}}\prod_{i}(\sigma_{i}^{*})^{i\Lambda_{i}}.
\end{equation}

The verification of the KLT relation for four points proceeds in two steps. The first consists of laying out the various ingredients in Mellin space, namely the gauge theory amplitudes and the KLT kernel. The second is comparing the Mellin space representation of the gravity amplitude, first computed in momentum space using the KLT relation to the gravity amplitude computed by applying the KLT relation to the Mellin amplitudes. It will be seen that the two quantities match, again up to an overall prefactor. Let us first note

\begin{equation}\label{eq:3.15}
    \mathcal{M}_{4,gauge}(1324) =  \frac{\Gamma\left[i\sum_{i}\Lambda_{i}\right]}{\left(i\sum_{i}\epsilon_{i}\sigma_{i}u_{i}\right)^{i\sum_{i}\Lambda_{i}}}C_{4}R_{4}\frac{\delta^{8}\left(C(\sigma^{*})\right)}{\sigma^{*}_{1}\sigma^{*}_{2}\sigma^{*}_{3}\sigma^{*}_{4}z_{13}z_{32}z_{24}z_{41}}\prod_{i}(\sigma_{i}^{*})^{i\Lambda_{i}} 
\end{equation}
and

\begin{equation}\label{eq:3.16}
    \mathcal{M}_{4,gauge}(1243) =  \frac{\Gamma\left[i\sum_{i}\Lambda_{i}\right]}{\left(i\sum_{i}\epsilon_{i}\sigma_{i}u_{i}\right)^{i\sum_{i}\Lambda_{i}}}C_{4}R_{4}\frac{\delta^{8}\left(C(\sigma^{*})\right)}{\sigma^{*}_{1}\sigma^{*}_{2}\sigma^{*}_{3}\sigma^{*}_{4}z_{12}z_{24}z_{43}z_{31}}\prod_{i}(\sigma_{i}^{*})^{i\Lambda_{i}}. 
\end{equation}

Computation of the four-graviton amplitude requires the relevant KLT kernels, which are the inverses of the objects,

\begin{equation}
    m(1234|1324)^{-1} = -\frac{\delta^{4}\left(\sum_{i}p_{i}\right)}{s_{14}}
\end{equation}
and

\begin{equation}
    m(1234|1243)^{-1} = -\frac{\delta^{4}\left(\sum_{i}p_{i}\right)}{s_{12}}
\end{equation}
respectively. The delta functions have been retained to make well defined the Mellin transforms,

\begin{equation}
    \mathcal{K}_{4}(1234|1324) = \int \prod_{i}\omega^{i\Lambda_{i}}_{i} m(1234|1324)^{-1}e^{i\sum_{i}\epsilon_{i}\omega_{i}u_{i}}d\omega_{i}
\end{equation}
and

\begin{equation}
    \mathcal{K}_{4}(1234|1243) = \int \prod_{i}\omega^{i\Lambda_{i}}_{i} m(1234|1243)^{-1}e^{i\sum_{i}\epsilon_{i}\omega_{i}u_{i}}d\omega_{i}.
\end{equation}
To evaluate these integrals, the critical observation is to realize that both integrands scale as $s^{-6}$ after changing to simplex variables. The rest of the integration process remains unchanged, giving

\begin{equation}\label{eq:3.21}
    \mathcal{K}_{4}(1234|1324) = \frac{\Gamma\left[-2+i\sum_{i}\Lambda_{i}\right]}{\left(i\sum_{i}\epsilon_{i}\sigma_{i}u_{i}\right)^{2+i\sum_{i}\Lambda_{i}}}C_{4}R_{4}\frac{1}{2\epsilon_{1}\epsilon_{4}\sigma^{*}_{1}\sigma^{*}_{4}z_{14}\z_{14}}\prod_{i}(\sigma_{i}^{*})^{i\Lambda_{i}}
\end{equation}
and

\begin{equation}\label{eq:3.22}
    \mathcal{K}_{4}(1234|1243) = \frac{\Gamma\left[-2+i\sum_{i}\Lambda_{i}\right]}{\left(i\sum_{i}\epsilon_{i}\sigma_{i}u_{i}\right)^{2+i\sum_{i}\Lambda_{i}}}C_{4}R_{4}\frac{1}{2\epsilon_{1}\epsilon_{2}\sigma^{*}_{1}\sigma^{*}_{2}z_{12}\z_{12}}\prod_{i}(\sigma_{i}^{*})^{i\Lambda_{i}}.
\end{equation}

A comparison of the Mellin space KLT relations to the momentum space relations can only be made once the gravity amplitudes in momentum space have been transformed. We first consider the gravity amplitude given by the KLT relation

\begin{equation}\label{eq:3.23}
    \A_{4}\left(1234\right) = \A_{4}\left(1234\right) m(1234|1324)\A_{4}\left(1324\right).
\end{equation}
where it is to be understood that we have a single momentum conserving delta function. 

Despite appearances, evaluation of the Mellin transform of the above is as straightforward as the integrals so far. The only difference here is the scaling with $s$ post conversion to simplex variables. The scaling here goes as $s^{-4-4+6} = s^{-2}$. The $s$ integral then becomes

\begin{equation}
   \int s^{4-1+i\sum_{i}\Lambda_{i} - 2}e^{is\sum_{i}\epsilon_{i}\sigma_{i}u_{i}}ds  = \frac{\Gamma\left[2+i\sum_{i}\Lambda_{i}\right]}{\left(i\sum_{i}\epsilon_{i}\sigma_{i}u_{i}\right)^{-2+i\sum_{i}\Lambda_{i}}}.
\end{equation}
It's now a simple matter to evaluate the simplex integrals by localization, giving

\begin{equation}\label{eq:3.25}
\begin{aligned}
    \mathcal{M}_{4,grav} =& \frac{\Gamma\left[2+i\sum_{i}\Lambda_{i}\right]}{\left(i\sum_{i}\epsilon_{i}\sigma_{i}u_{i}\right)^{-2+i\sum_{i}\Lambda_{i}}}R_{4}C_{4}\times\delta^{16}\left(C(\sigma^{*})\right)\times\\
    &\frac{2\epsilon_{1}\epsilon_{4}\sigma^{*}_{1}\sigma^{*}_{4}z_{14}\z_{14}}{(\sigma^{*}_{1}\sigma^{*}_{2}\sigma^{*}_{3}\sigma^{*}_{4}z_{12}z_{23}z_{34}z_{41})(\sigma^{*}_{1}\sigma^{*}_{2}\sigma^{*}_{3}\sigma^{*}_{4}z_{13}z_{32}z_{24}z_{41})}\prod_{i}(\sigma_{i}^{*})^{i\Lambda_{i}}.
\end{aligned}
\end{equation}
It is a simple matter to verify from (\emph{\ref{eq:3.14}}), (\emph{\ref{eq:3.15}}) and (\emph{\ref{eq:3.21}}) that

\begin{equation}\label{eq:3.26}
    (\emph{\ref{eq:3.25}}) = \mathcal{M}_{4,gauge}(1234)\Delta_{4}\mathcal{K}(1234|1324)^{-1}\mathcal{M}_{4,gauge}(1324)
\end{equation}
where

\begin{equation}
    \Delta_{4} = \frac{\Gamma\left[-2+i\sum_{i}\Lambda_{i}\right]\Gamma\left[2+i\sum_{i}\Lambda_{i}\right]}{\Gamma\left[i\sum_{i}\Lambda_{i}\right]^{2}}.
\end{equation}
The relation contained in (\emph{\ref{eq:3.26}}) is the Mellin space avatar of the kinematic KLT relation (\emph{\ref{eq:3.23}}). We see again that the KLT relation continues to hold up to a prefactor that has no kinematical content. 

To carry out one more test of the KLT relations, we take the four-graviton amplitude defined according to

\begin{equation}\label{eq:3.28}
    \A_{4}\left(1234\right) = \A_{4}\left(1234\right) m(1234|1243)\A_{4}\left(1243\right).
\end{equation}
It is a short exercise to verify that this is indeed equal to (\emph{\ref{eq:3.23}}). Following the same procedure, it is clear that we have this time

\begin{equation}\label{eq:3.29}
\begin{aligned}
    \mathcal{M}_{4,grav} =& \frac{\Gamma\left[-2+i\sum_{i}\Lambda_{i}\right]}{\left(i\sum_{i}\epsilon_{i}\sigma_{i}u_{i}\right)^{-2+i\sum_{i}\Lambda_{i}}}R_{4}C_{4}\times\delta^{16}\left(C(\sigma^{*})\right)\times\\
    &\frac{2\epsilon_{1}\epsilon_{2}\sigma^{*}_{1}\sigma^{*}_{2}z_{12}\z_{12}}{(\sigma^{*}_{1}\sigma^{*}_{2}\sigma^{*}_{3}\sigma^{*}_{4}z_{12}z_{23}z_{34}z_{41})(\sigma^{*}_{1}\sigma^{*}_{2}\sigma^{*}_{3}\sigma^{*}_{4}z_{12}z_{24}z_{43}z_{41})}\prod_{i}(\sigma_{i}^{*})^{i\Lambda_{i}}.
\end{aligned}
\end{equation}
It follows now from equations (\emph{\ref{eq:3.14}}), (\emph{\ref{eq:3.16}}) and (\emph{\ref{eq:3.22}}) that once again

\begin{equation}\label{eq:3.30}
    (\emph{\ref{eq:3.29}}) = \mathcal{M}_{4,gauge}(1234)\Delta_{4}\mathcal{K}(1234|1243)^{-1}\mathcal{M}_{4,gauge}(1243)
\end{equation}
establishing again the persistence of the KLT relation in Mellin space for four-particle scattering. 

This calculations carried out in this section should convince the reader that the KLT relations continue to hold for the four particle case due to localization properties of the Mellin integral at this multiplicity. Drawing on this observation, we are led to expect that this should continue to be the case for five-particle scattering. To test this, we proceed to perform the same analysis for the five-particle case.

\subsection{Five-Particle Amplitudes}\label{ssec:3}
Our concern now will be the treatment of five-particle amplitudes in the $\mathcal{N}=4$ and $\mathcal{N}=8$ theories and a verification of the KLT relations between them. At tree level, there are two nontrivial helicity sectors - the MHV and anti-MHV sector. The latter however is related to the former via complex conjugation, allowing us to focus on the MHV case alone. Accordingly, we start with the amplitude given by

\begin{equation}\label{eq:4.1}
    \mathcal{A}_{5,gauge}(12345) = \frac{\delta^{8}\left(C(\lambda)\right)\delta^{4}\left(\sum_{i}p_{i}\right)}{\ab{12}\ab{23}\ab{34}\ab{45}\ab{51}}
\end{equation}
where we have denoted

\begin{equation}
    C(\lambda)_{\alpha}^{I} = \sum_{i=1}^{n}\lambda^{(i)}_{\alpha}\eta^{I}_{(i)}.
\end{equation}

Being interested in verifying the KLT relations, we will also have occasion to use the permuted amplitudes

\begin{equation}
    \mathcal{A}_{5,gauge}(12435) = \frac{\delta^{8}\left(C(\lambda)\right)\delta^{4}\left(\sum_{i}p_{i}\right)}{\ab{12}\ab{24}\ab{43}\ab{35}\ab{51}},
\end{equation}

\begin{equation}
    \mathcal{A}_{5,gauge}(13254) = \frac{\delta^{8}\left(C(\lambda)\right)\delta^{4}\left(\sum_{i}p_{i}\right)}{\ab{13}\ab{32}\ab{25}\ab{54}\ab{41}}
\end{equation}
and

\begin{equation}
    \mathcal{A}_{5,gauge}(14253) = \frac{\delta^{8}\left(C(\lambda)\right)\delta^{4}\left(\sum_{i}p_{i}\right)}{\ab{14}\ab{42}\ab{25}\ab{53}\ab{31}}.
\end{equation}

The first task at hand is now the evaluation of the Mellin transforms of the foregoing amplitudes. As always, we can perform the evaluation for the amplitude (\emph{\ref{eq:4.1}}) and infer the remaining ones by analogy as we did in the last section. The first step naturally is analyzing the $s$-scaling following the coordinate transformation,

\begin{equation}
    s = \sum_{i}^{5}\omega_{i},
\end{equation}

\begin{equation}
    \sigma_{i} = s^{-1}\omega_{i}.
\end{equation}

Simple inspection reveals the scaling,

\begin{equation}
    \mathcal{A}_{5,gauge} = s^{-5}\hat{\mathcal{A}}_{5,gauge}
\end{equation}
where the hat as usual indicates the amplitude $\mathcal{A}_{5,gauge}$ with every $\omega_{i}$ replaced by $\sigma_{i}$. This observation implies that the Mellin transform

\begin{equation}
    \mathcal{M}_{5,gauge}(12345) = \int \prod_{i}\omega^{i\Lambda_{i}}_{i} \A_{5}\left(12345\right)e^{i\sum_{i}\epsilon_{i}\omega_{i}u_{i}}d\omega_{i}
\end{equation}
becomes

\begin{equation}
    \mathcal{M}_{5,gauge}(12345) = \int s^{5-1-5} \prod_{i}\sigma^{i\Lambda_{i}}_{i} \hat{\A}_{5}\left(12345\right)e^{is\sum_{i}\epsilon_{i}\sigma_{i}u_{i}}d\sigma_{i}ds
\end{equation}
keeping in mind of course that the $\sigma_{i}$ are restricted to lie on the five-simplex. The $s$ integral is evaluated to give

\begin{equation}\label{eq:4.11}
    \mathcal{M}_{5,gauge}(12345) = \frac{\Gamma\left[i\sum_{i}\Lambda_{i}\right]}{\left(i\sum_{i}\epsilon_{i}\sigma^{*}_{i}u_{i}\right)^{i\sum_{i}\Lambda_{i}}} \int  \prod_{i}\sigma^{i\Lambda_{i}}_{i} \hat{\A}_{5}\left(12345\right)d\sigma_{i}.
\end{equation}
The simplex integrals are evaluated once again by localization via the replacement

\begin{equation}\label{eq:4.12}
    \delta^{4}\left(\sum_{i}\epsilon_{i}\sigma_{i}q_{i}\right) \delta\left(1-\sum_{i}\sigma_{i}\right) = J_{5}\prod_{i}\delta(\sigma_{i}-\sigma^{*}_{i})
\end{equation}
To spare the reader the view of the $\sigma^{*}_{i}$ their values have been provided in full in an attached \textsc{Mathematica} notebook. $J_{5}$ is a Jacobian which has also been included in the notebook. That being said, we can now effect the localizations, which leave us with

\begin{equation}
    \mathcal{M}_{5,gauge}(12345) = \frac{\Gamma\left[i\sum_{i}\Lambda_{i}\right]}{\left(\sum_{i}\epsilon_{i}\sigma^{*}_{i}u_{i}\right)^{i\sum_{i}\Lambda_{i}}} J_{5}\frac{\delta^{8}\left(C(\sigma^{*})\right)}{\sigma^{*}_{1}\sigma^{*}_{2}\sigma^{*}_{3}\sigma^{*}_{4}\sigma^{*}_{5}z_{12}z_{23}z_{34}z_{45}z_{51}}\prod_{i}(\sigma_{i}^{*})^{i\Lambda_{i}}.
\end{equation}
The Mellin transforms of the remaining relevant gauge theory amplitudes are evaluated similarly - by analogy they become

\begin{equation}
    \mathcal{M}_{5,gauge}(12435) = \frac{\Gamma\left[i\sum_{i}\Lambda_{i}\right]}{\left(i\sum_{i}\epsilon_{i}\sigma^{*}_{i}u_{i}\right)^{i\sum_{i}\Lambda_{i}}} J_{5}\frac{\delta^{8}\left(C(\sigma^{*})\right)}{\sigma^{*}_{1}\sigma^{*}_{2}\sigma^{*}_{3}\sigma^{*}_{4}\sigma^{*}_{5}z_{12}z_{24}z_{43}z_{35}z_{51}}\prod_{i}(\sigma_{i}^{*})^{i\Lambda_{i}},
\end{equation}

\begin{equation}
    \mathcal{M}_{5,gauge}(13254) = \frac{\Gamma\left[i\sum_{i}\Lambda_{i}\right]}{\left(i\sum_{i}\epsilon_{i}\sigma^{*}_{i}u_{i}\right)^{i\sum_{i}\Lambda_{i}}} J_{5}\frac{\delta^{8}\left(C(\sigma^{*})\right)}{\sigma^{*}_{1}\sigma^{*}_{2}\sigma^{*}_{3}\sigma^{*}_{4}\sigma^{*}_{5}z_{13}z_{32}z_{25}z_{54}z_{41}}\prod_{i}(\sigma_{i}^{*})^{i\Lambda_{i}},
\end{equation}
and

\begin{equation}
    \mathcal{M}_{5,gauge}(14253) = \frac{\Gamma\left[i\sum_{i}\Lambda_{i}\right]}{\left(i\sum_{i}\epsilon_{i}\sigma^{*}_{i}u_{i}\right)^{i\sum_{i}\Lambda_{i}}} J_{5}\frac{\delta^{8}\left(C(\sigma^{*})\right)}{\sigma^{*}_{1}\sigma^{*}_{2}\sigma^{*}_{3}\sigma^{*}_{4}\sigma^{*}_{5}z_{14}z_{42}z_{25}z_{53}z_{31}}\prod_{i}(\sigma_{i}^{*})^{i\Lambda_{i}}.
\end{equation}

The computation of the five graviton amplitude in the MHV sector is evaluated by application of the five-particle KLT relation. Specifically, we have the expression \cite{Mizera:2016jhj},

\begin{equation}
\begin{aligned}
    \mathcal{A}_{5,grav} =& m(12345|13254)\A_{5,gauge}(12345)\A_{5,gauge}(13254)\\& +  m(12435|14253)\A_{5,gauge}(12435)\A_{5,gauge}(14253)
\end{aligned}
\end{equation}
where

\begin{equation}
    m(12345|13254)^{-1} = \frac{\delta^{4}\left(\sum_{i}p_{i}\right)}{s_{23}s_{45}},
\end{equation}
and

\begin{equation}
    m(12435|14253)^{-1} = \frac{\delta^{4}\left(\sum_{i}p_{i}\right)}{s_{24}s_{35}}
\end{equation}
Our task now is to convince ourselves that this formula can be applied minimally modified in Mellin space as well. Indeed, five particles is the maximal number at which the localization formula can be applied. Accordingly, it should not be surprising that the KLT relation continues to be true. 

We start by performing the modified Mellin transform to the last formula. Due to the rather lengthy nature of the expressions, we will rewrite the modified transform as

\begin{equation}
    \mathcal{M}_{5,grav} = \mathcal{M}^{(1)}_{5,grav} + \mathcal{M}^{(2)}_{5,grav}
\end{equation}
where

\begin{equation}\label{eq:4.21}
 \mathcal{M}^{(1)}_{5,grav}=   \int \prod_{i}\omega^{i\Lambda_{i}}_{i} \left(m(12345|13254)\A_{5,gauge}(12345)\A_{5,gauge}(13254)\right)e^{i\sum_{i}\epsilon_{i}\omega_{i}u_{i}}d\omega_{i},
\end{equation}
and

\begin{equation}\label{eq:4.22}
    \mathcal{M}^{(2)}_{5,grav}=   \int \prod_{i}\omega^{i\Lambda_{i}}_{i} \left(m(12435|14253)\A_{5,gauge}(12435)\A_{5,gauge}(14253)\right)e^{i\sum_{i}\epsilon_{i}\omega_{i}u_{i}}d\omega_{i}.
\end{equation}
Since there is little qualitative difference between these two integrals, we can compute the first and infer the second by inspection. We first note that following scaling with $s$ after transformation to simplex variables

\begin{equation}
    \mathcal{A}_{5,grav} = s^{-5-5+8}\hat{\A}_{5,grav}.
\end{equation}
The $s$ integral post variable change is carried out by then noting,

\begin{equation}
    \int s^{5-1-2}e^{is\sum_{i}\sigma_{i}u_{i}} ds = \frac{\Gamma\left[3+i\sum_{i}\Lambda_{i}\right]}{i\left(\sum_{i}\epsilon_{i}\sigma^{*}_{i}u_{i}\right)^{3+i\sum_{i}\Lambda_{i}}}.
\end{equation}
Applying the residue theorem according to the prescription of equation (\emph{\ref{eq:4.12}}), the integral in (\emph{\ref{eq:4.21}}) is

\begin{equation}
\begin{aligned}
\mathcal{M}_{5,grav}^{(1)} = & \frac{\Gamma\left[3+i\sum_{i}\Lambda_{i}\right]}{\left(\sum_{i}\sigma^{*}_{i}u_{i}\right)^{3+i\sum_{i}\Lambda_{i}}} J_{5}\frac{\delta^{16}\left(C(\sigma^{*})\right)}{\sigma^{*}_{1}\sigma^{*}_{2}\sigma^{*}_{3}\sigma^{*}_{4}\sigma^{*}_{5}z_{12}z_{23}z_{34}z_{45}z_{51}}\\
&\times\frac{4\epsilon_{2}\epsilon_{3}\epsilon_{4}\epsilon_{5}\sigma^{*}_{2}\sigma^{*}_{3}\sigma^{*}_{4}\sigma^{*}_{5}z_{23}\z_{23}z_{45}\z_{45}}{\sigma^{*}_{1}\sigma^{*}_{2}\sigma^{*}_{3}\sigma^{*}_{4}\sigma^{*}_{5}z_{13}z_{32}z_{25}z_{54}z_{51}}\prod_{i}(\sigma_{i}^{*})^{i\Lambda_{i}}.
\end{aligned}
\end{equation}
Analogously, the transform in (\emph{\ref{eq:4.22}}) is found to be

\begin{equation}
\begin{aligned}
\mathcal{M}_{5,grav}^{(2)} = & \frac{\Gamma\left[3+i\sum_{i}\Lambda_{i}\right]}{\left(\sum_{i}\sigma^{*}_{i}u_{i}\right)^{3+i\sum_{i}\Lambda_{i}}} J_{5}\frac{\delta^{16}\left(C(\sigma^{*})\right)}{\sigma^{*}_{1}\sigma^{*}_{2}\sigma^{*}_{3}\sigma^{*}_{4}\sigma^{*}_{5}z_{12}z_{24}z_{43}z_{35}z_{51}}\\
&\times\frac{4\epsilon_{2}\epsilon_{3}\epsilon_{4}\epsilon_{5}\sigma^{*}_{2}\sigma^{*}_{3}\sigma^{*}_{4}\sigma^{*}_{5}z_{24}\z_{24}z_{35}\z_{35}}{\sigma^{*}_{1}\sigma^{*}_{2}\sigma^{*}_{3}\sigma^{*}_{4}\sigma^{*}_{5}z_{14}z_{42}z_{25}z_{53}z_{31}}\prod_{i}(\sigma_{i}^{*})^{i\Lambda_{i}}.
\end{aligned}
\end{equation}
Checking the KLT kernel now requires that we evaluate the Mellin transforms of the inverse kernels 

\begin{equation}
   \mathcal{K}(12345|13254)=   \int \prod_{i}\omega^{i\Lambda_{i}}_{i} m(12345|13254)^{-1}e^{i\sum_{i}\epsilon_{i}\omega_{i}u_{i}}d\omega_{i},
\end{equation}
and

\begin{equation}
    \mathcal{K}(12435|14253)=   \int \prod_{i}\omega^{i\Lambda_{i}}_{i} m(12435|14253)^{-1}e^{i\sum_{i}\epsilon_{i}\omega_{i}u_{i}}d\omega_{i}.
\end{equation}
Once again, all we need is the scaling with $s$, after which the simplex integrals are found by the residue prescription. We have the scaling $s^{-8}$ with $s$, which can be checked by simple inspection. Accordingly, the $s$ integral for both of the above transforms becomes,

\begin{equation}
    \int s^{5-1-8}e^{is\sum_{i}\epsilon_{i}\sigma_{i}u_{i}}ds = \frac{\Gamma\left[-3+i\sum_{i}\Lambda_{i}\right]}{\left(i\sum_{i}\epsilon_{i}\sigma^{*}_{i}u_{i}\right)^{-3+i\sum_{i}\Lambda_{i}}}.
\end{equation}

The simplex integrals once evaluated give us the simple formulae

\begin{equation}
    \mathcal{K}(12345|13254) = \frac{\Gamma\left[-3+i\sum_{i}\Lambda_{i}\right]}{\left(i\sum_{i}\epsilon_{i}\sigma^{*}_{i}u_{i}\right)^{-3+i\sum_{i}\Lambda_{i}}}\frac{J_{5}}{4\epsilon_{2}\epsilon_{3}\epsilon_{4}\epsilon_{5}\sigma^{*}_{2}\sigma^{*}_{3}\sigma^{*}_{4}\sigma^{*}_{5}z_{23}\z_{23}z_{45}\z_{45}}\prod_{i}(\sigma_{i}^{*})^{i\Lambda_{i}}
\end{equation}
and

\begin{equation}
    \mathcal{K}(12435|14253) = \frac{\Gamma\left[-3+i\sum_{i}\Lambda_{i}\right]}{\left(\sum_{i}\sigma^{*}_{i}u_{i}\right)^{-3+i\sum_{i}\Lambda_{i}}}\frac{J_{5}}{4\epsilon_{2}\epsilon_{3}\epsilon_{4}\epsilon_{5}\sigma^{*}_{2}\sigma^{*}_{3}\sigma^{*}_{4}\sigma^{*}_{5}z_{24}\z_{24}z_{35}\z_{35}}\prod_{i}(\sigma_{i}^{*})^{i\Lambda_{i}}.
\end{equation}
The KLT relation is verified by noting the equalities

\begin{equation}
    \mathcal{M}_{5,grav}^{(1)} =  \mathcal{M}_{5,gauge}(12345)\Delta_{5}\mathcal{K}(12345|13254)^{-1}\mathcal{M}_{5,gauge}(13254),
\end{equation}
and

\begin{equation}
    \mathcal{M}_{5,grav}^{(2)} =  \mathcal{M}_{5,gauge}(12435)\Delta_{5}\mathcal{K}(12435|14253)^{-1}\mathcal{M}_{5,gauge}(14253)
\end{equation}
where

\begin{equation}
    \Delta_{5} = \frac{\Gamma\left[-3+i\sum_{i}\Lambda_{i}\right]\Gamma\left[3+i\sum_{i}\Lambda_{i}\right]}{\Gamma\left[i\sum_{i}\Lambda_{i}\right]^{2}}.
\end{equation}

\section{Obstructions at Higher Multiplicity}\label{sec:3}
Having established the working of the KLT relations for low multiplicity, we can now discuss why this works, as well as the obstructions at higher multiplicity. To do this, we first consider a related problem, namely that of supersymmetry\footnote{The discussion of preserving supersymmetry in this section was heavily informed by conversations I have had with Jacob Bourjaily, Cameron Langer and Kokkimidis Patatoukos, for which I am grateful.}.

Consider first the form of a scattering amplitude in a supersymmetric theory. 

\begin{equation}
 \mathcal{A} =   f\left(\lbrace{\lambda_{i},\tilde{\lambda}_{i}\rbrace}\right)\delta^{2\mathcal{N}}\left(C(\lambda)\right)
\end{equation}
where the matrix $C$ is some kinematic function linear in the Grassmann variables generating supersymmetry. While we are restricted in this note to the study of scattering amplitudes in maximally supersymmetric theories, but the arguments we will discuss hold quite generally. 

Noting that such a representation exists in momentum space, we now ask under what conditions we expect it to persist, as least qualitatively, in some other basis. Although we have considered the Mellin basis in this note, the question is worth pondering for any basis change. Suppose the basis change is effected by a residue prescription. To keep things general, we denote the transformed variables by $\lbrace{x\rbrace}$ and the new variables by $\lbrace{y\rbrace}$. In the case  of the Mellin transform, the original variables would be the simplex variables, and the new ones would be the simplex coordinates after the delta functions have been transformed. At any rate, the change of basis is provided by a residue prescription,

\begin{equation}
    \oint_{|x-y|<\epsilon}\mathcal{A} = f\left(\lbrace{\lambda_{i},\tilde{\lambda}_{i}\rbrace}\right)\delta^{2\mathcal{N}}\left(C(\lambda)\right)|_{x=y}.
\end{equation}
Written this way, it is clear that a residue theorem applied on the amplitude in supersymmetric form does nothing to damage the supersymmetric representation. There is no reason however to believe that a manifest supersymmetric expression continues to apply if more complicated integrals are involved. 

This brings us to the amplitudes of interest to us in this paper. Indeed it is due in large part to this last comment that we have restricted attention to amplitudes at low multiplicity. We saw that for amplitudes with up to five external particles the delta function constraints enforcing momentum conservation and the simplex property localized the transform on special values of the simplex variables, which could be found explicitly by solving the delta function constraints. The complication at higher multiplicity arises due to the fact that the solutions $\sigma^{*}_{i}$ for $i\leq 5$ will in general be functions of the remaining simplex variables $\sigma_{i\geq 6}$. Schematically, the regulated Mellin transform will behave as

\begin{equation}
\begin{aligned}
    \int &\frac{\Gamma\left[i\sum_{i}\Lambda_{i}\right]}{\left(i\sum_{i\leq 5}\epsilon_{i}u_{i}\sigma^{*}_{i}+i\sum_{i\geq 6}\epsilon_{i}u_{i}\sigma_{i}\right)^{(n-2)+i\sum_{i}\Lambda_{i}}}\times\\&\hat{\mathcal{A}}_{n,gauge}\left(\lbrace{\sigma^{*}_{i\leq 5}(\sigma_{i\geq 6})\rbrace},\lbrace{\sigma_{i\geq 6}\rbrace}\right)\prod _{i\leq 5}\left(\sigma_{i}^{*}\right)^{i\Lambda_{i}}\prod_{i\geq 6}\sigma_{i}^{\Lambda_{i}}d\sigma_{i}
\end{aligned}
\end{equation}
for gauge theory amplitudes and

\begin{equation}
\begin{aligned}
    \int& \frac{\Gamma\left[(n-2)+i\sum_{i}\Lambda_{i}\right]}{\left(i\sum_{i\leq 5}\epsilon_{i}u_{i}\sigma^{*}_{i}+i\sum_{i\geq 6}\epsilon_{i}u_{i}\sigma_{i}\right)^{(n-2)+i\sum_{i}\Lambda_{i}}}\times\\&\hat{\mathcal{A}}_{n,gravity}\left(\lbrace{\sigma^{*}_{i\leq 5}(\sigma_{i\geq 6})\rbrace},\lbrace{\sigma_{i\geq 6}\rbrace}\right)\prod _{i\leq 5}\left(\sigma_{i}^{*}\right)^{i\Lambda_{i}}\prod_{i\geq 6}\sigma_{i}^{\Lambda_{i}}d\sigma_{i}
\end{aligned}
\end{equation}
for gravity amplitudes. Needless to say, there is no obvious sense in which manifest supersymmetry can be guaranteed. The amplitude however is still a linear combination of quantities exhibiting manifest supersymmetry.

It is on account of precisely the same facts that the KLT relations are valid when (more or less) blindly applied only until we move past five-particle scattering. The KLT relation is a nonlinear relation between gauge theory and gravity amplitudes at tree level. However, even a relation like this will no doubt continue to hold when a residue theorem is applied on them. Even beyond five particles however, we do have some basic facts to salvage. In particular, suppose we work with two BCJ bases $\alpha$ and $\beta$ each of length $(n-3)!$ and take the KLT relation in the form \cite{Mizera:2016jhj}

\begin{equation}
    \A_{n,gravity} = \sum_{\alpha,\beta}\A_{n,gauge}[\alpha]K[\alpha|\beta]^{-1}\A_{n,gauge}[\beta].
\end{equation}
where the sum is carried out over all elements of the bases $\alpha$ and $\beta$, which are permutations over the external particles. Although this will not continue to hold at the level of the full Mellin transform, they will continue to hold at a partial integrand level, namely after the first five simplex variables have been integrated out. Schematically speaking we have,

\begin{equation}
\begin{aligned}
    & \mathcal{M}_{n,gravity}(\lbrace{\sigma^{*}_{i\leq 5}\rbrace},\lbrace{\sigma_{i\geq 6}\rbrace}) =\\ &\sum_{\alpha,\beta}\mathcal{M}_{n,gauge}[\alpha](\lbrace{\sigma^{*}_{i\leq 5}\rbrace},\lbrace{\sigma_{i\geq 6}\rbrace})\Delta_n\mathcal{K}[\alpha|\beta]^{-1}(\lbrace{\sigma^{*}_{i\leq 5}\rbrace},\lbrace{\sigma_{i\geq 6}\rbrace})\mathcal{M}_{n,gauge}[\beta](\lbrace{\sigma^{*}_{i\leq 5}\rbrace},\lbrace{\sigma_{i\geq 6}\rbrace}).
\end{aligned}
\end{equation}
where it is to be understood that the $s$ integrals have been carried out, along with the first five simplex integrals as well. Now, the inverse KLT kernel $K^{-1}$ must be corrected by a factor $\Delta_n$ given by

\begin{equation}\label{eq:5.7}
    \Delta_{n} = \frac{\Gamma\left[-(n-2)+i\sum_{i}\Lambda_{i}\right]\Gamma\left[(n-2)+i\sum_{i}\Lambda_{i}\right]}{\Gamma\left[i\sum_{i}\Lambda_{i}\right]^{2}}.
\end{equation}
Although we have been somewhat telegraphic, the discussions of the last three sections give this assertion some context. The KLT relations continue to hold so long as we work only with residues, which is the case up till five simplex integrals are carried out. The correction factor is multiplicative, and is given by (\emph{\ref{eq:5.7}}).

\section{Conclusions}\label{sec:4}
There are two problems that we have considered in this article - those of preserving manifest supersymmetry and implementing the KLT relations among gauge theory and gravity amplitudes in the celestial basis. Indeed, we found that these two problems were closely related. Celestial amplitudes could be written expressing manifest supersymmetry and satisfy the KLT relations under the same condition, namely that the change of basis could be regarded as a residue prescription. 

A natural step forward would be to find a genuine analogue of the KLT relations in Mellin space. While we provided an integrand level analogue in section \ref{sec:4}, the work involved in obtaining a full-fledged KLT relation or finding that one does not exist might be instructive.

One possible route towards exploring this problem would be to consider the framework laid out recently in \cite{Casali:2020vuy,Casali:2020uvr}. In the latter work, the authors have framed the Mellin space amplitudes for gauge theory and gravity in terms of a twisted cohomology theory (see \cite{Mizera:2017cqs,Mizera:2017rqa,Mizera:2019blq,Mizera:2019gea,Mizera:2020wdt,Mastrolia:2018uzb,Frellesvig:2019kgj,Frellesvig:2019uqt} for details). Twisted cohomology has the added advantage of trivializing the double copy as a statement about linear algebra. Accordingly, this may be one systematic way to proceed. At the same time, it is worth pointing out that in this work, it was the unregulated Mellin transform under consideration. There will be some work to do in order to generalize to the regulated form. 

One concrete direction for future research is the development and study of recursion relations \cite{Britto:2004ap,Britto:2005fq} in the case of Mellin amplitudes. Direct evaluation of Mellin amplitudes for high multiplicity generally gives rise to rather complicated objects \cite{Schreiber:2017jsr}, so one is led to hope that there may be a simpler approach to computing these quantities. In particular, the expression of recursion in terms of on-shell diagrams \cite{ArkaniHamed:2010kv,ArkaniHamed:2012nw} and the associated relationship to the Grassmannian may prove helpful in this regard.

It should also be noted that the motion into Mellin space might be relevant to the broader goal of understanding potential underlying geometries and worldsheet formulations of quantum field theories, at least at tree level. While scalar theories generically give rise to complicated geometries \cite{Arkani-Hamed:2017mur,Banerjee:2018fgd,Raman:2019utu,Jagadale:2019byr,Kalyanapuram:2019nnf,Kalyanapuram:2020axt,Kalyanapuram:2020tsr,Kalyanapuram:2020vil}, it is often easier and more natural to establish holographic structures for gauge theory and gravity \cite{Adamo:2017nia,Adamo:2018mpq,Kalyanapuram:2020epb}. In light of such dual pictures and the analytic control we are gaining over the Mellin space picture, a synthesis of these ideas may be feasible, manifesting double copy relations like BCJ and KLT more effectively.

\section*{Acknowledgements} 
I am indebted to Jacob Bourjaily for going over the draft, suggesting a number of improvements and his persistent encouragement. I am grateful to Shamik Banerjee, Eduardo Casali, Sudip Ghosh and Monica Pate for comments on the draft. A special thanks goes to Arnab Priya Saha for providing his computations on four-particle amplitudes, which inspired the present investigation. I also thank Jacob Bourjaily, Cameron Langer and Kokkimidis Patatoukos for discussions on an ongoing collaboration, the details of which helped clarify conceptual points of the present work. This project has been supported by an ERC Starting Grant (No. 757978) and a grant from the Villum Fonden (No. 15369).


\bibliographystyle{utphys}
\bibliography{v1.bib}
\end{document}